\documentclass[sigconf]{acmart}
\AtBeginDocument{%
  }

\copyrightyear{2025}
\acmYear{2025}
\acmConference[XAIxArts 2025]{Explainable AI for the Arts Workshop 2025}{June 23, 2025}{Online} 

\usepackage{url}
\usepackage{hyperref}
\usepackage{float}
\usepackage{graphicx}
\usepackage{subfigure}
\usepackage{subcaption} 
\usepackage{caption} 
\usepackage{comment}

\begin{document}


\title{Explainability-in-Action: Enabling Expressive Manipulation and Tacit Understanding by Bending Diffusion Models in ComfyUI}


\author{Ahmed M. Abuzuraiq}
\affiliation{%
\department{School of Interactive Arts and Technology}
  \institution{Simon Fraser University}
  \city{Surrey}
  \country{Canada}}
  \orcid{https://orcid.org/0000-0002-3604-7623}
\email{aabuzura@sfu.ca}

\author{Philippe Pasquier}
\affiliation{%
  \department{School of Interactive Arts and Technology}
  \institution{Simon Fraser University}
  \city{Surrey}
  \country{Canada}}
  \orcid{https://orcid.org/0000-0001-8675-3561}
\email{pasquier@sfu.ca}

\renewcommand{\shortauthors}{Abuzuraiq and Pasquier}

\begin{abstract}
  Explainable AI (XAI) in creative contexts can go beyond transparency to support artistic engagement, modifiability, and sustained practice. While curated datasets and training human-scale models can offer artists greater agency and control, large-scale generative models like text-to-image diffusion systems often obscure these possibilities. We suggest that even large models can be treated as creative materials if their internal structure is exposed and manipulable. We propose a craft-based approach to explainability rooted in long-term, hands-on engagement—akin to Schön’s "reflection-in-action"—and demonstrate its application through a model-bending and inspection plugin integrated into the node-based interface of ComfyUI. We demonstrate that by interactively manipulating different parts of a generative model, artists can develop an intuition about how each component influences the output.
\end{abstract}

\keywords{Explainable AI,
Generative AI,
Diffusion Models,
Visual Art,
Model Bending,
Model Crafting,
Reflection-in-Action}


\maketitle

\section{Introduction}
Explainable AI (XAI) traditionally focuses on demystifying machine learning systems for auditing, transparency, or safety~\cite{gunning2019xai}. However, explainability in creative contexts can serve different roles, including making models modifiable and debuggable for a meaningful artistic engagement~\cite{bryan2023exploring}, and creating the conditions for a sustained artistic practice that goes beyond mere amusement~\cite{tecks_2024_ExplainabilityPathsSustained}. The last workshop on XAIxArts~\cite{bryan-kinns_2025_XAIxArtsManifestoExplainablea} culminated in a manifesto that called artists and researchers to explore alternatives to the technocentric explanations of AI that value artistic practices and use intentional hacking, glitches, and imperfections as creative tools. This work continues along those lines.

It is argued that working with curated (small) datasets and human-scale models can enhance artists’ agency~\cite{vigliensoni_2022_smalldatamindsetgenerative}. Broadening the scope of the artistic process with AI to include both model training and inference can also reinforce artists' trust in AI models~\cite{tecks_2024_ExplainabilityPathsSustained}. Both approaches afford artists more control over AI models. However, artists' ability to craft with AI models as materials for creation~\cite{caramiaux2022explorers, broad_2024_using, rozental_2025_HowArtistsUse} is diminished when working with large-scale generative models~\cite{abuzuraiq_2024_SeizingMeansProduction}. If small-data and model training present an alternate explainable way for artists working with human-scale generative models, what options are available for artists wanting to work with large-scale models such as text-to-image diffusion models? Although this material relationship is easier to build with small models, where data and model structure/training are accessible and malleable, it becomes more challenging at scale. However, we argue that large models can also be treated as material, provided their structure is exposed and can be manipulated.

We propose that treating AI models as materials, especially within interfaces that expose the model's components like the node-based interface of ComfyUI, can foster a craft-based relationship between artists and generative systems. In particular, through long-term engagement and hands-on manipulation of a model's components, artists can develop a form of tacit understanding and familiarity with their AI materials akin to that found in traditional crafts, i.e. a form of explainability that is rooted in doing, aligned with Schön’s "reflection-in-action"~\cite{schon1983reflective}. We explore the application of these ideas for large-scale models through a plugin for model bending~\cite{broad_2021_NetworkbendingExpressive} that is implemented into the node-based interface of ComfyUI~\cite{2024_ComfyAI}. 

\section{From Models as Commodity to Models as Material}
Model repositories like CivitAI~\cite{2025_civitai} and TensorArt~\cite{2025_tensorart} provide users with thousands of pre-trained and fine-tuned models, generated images with associated prompts and parameters, and reproducible workflows for a wide variety of use-cases and styles. The quantity and diversity of models and workflows shared online are simply unprecedented within the context of generative art, and cloud computing services are making them more accessible. A large portion of these models are adaptations and personalizations (e.g. Low-Rank Adaptations -- LoRAs~\cite{hu_2021_LoraLowrankadaptation}) of a smaller set of base models (e.g. Stable Diffusion~\cite{stabilityai_2024_stablediffusion} and Flux variants). Following Abonamah et al.~\cite{abonamah2021commoditization}, we argue that the proliferation of generative AI models will likely result in their commodification — a development that, we contend, may further perpetuate their biases~\cite{vazquez2024taxonomy}.

When technology becomes commodified, its design priorities shift—from serving as a raw material for creation (as with early computer terminals) to emphasizing accessibility and replaceability. This shift can bring innovation benefits, as seen with personal computers. However, in the case of generative AI, commodification is occurring at an unprecedented pace. The challenge, however, is that this rapid commodification may discourage sustained engagement with individual models. Instead of fostering deep familiarity with their inner workings, affordances, and limitations, users are incentivized to sample from an ever-expanding catalogue without developing the tacit knowledge necessary for critical or innovative use. This dynamic may also contribute to \textit{AI fatigue}—a feeling of exhaustion or overwhelm, potentially driven by the rapid pace of updates to AI models and tools, as well as the opaqueness and complexity of AI systems~\footnote{\url{https://newsletter.victordibia.com/p/you-have-ai-fatigue-thats-why-you}}. Historically, artists have cultivated a deep, material understanding of their tools—whether brushes, cameras, or software—as a prerequisite for meaningful creative expression. Generative models require the same kind of sustained engagement to be used critically and responsibly. At the same time, many artists—who are uniquely positioned to critique and subvert these systems~\cite{bryan-kinns_2025_XAIxArtsManifestoExplainablea, broad_2024_using}—are choosing not to engage with them on principle, citing concerns that these models are trained on the work of unattributed artists~\cite{kawakami2024impact}. Therefore, we argue that it's imperative to encourage and offer tools for artists to engage with AI in ethical~\footnote{The CommonCanvas~\cite{gokaslan2024commoncanvas}, trained on Creative Commons images, may offer an alternative for artists who find using large generative models questionable.}, sustained~\cite{tecks_2024_ExplainabilityPathsSustained} and critical~\cite{broad_2024_using} ways, and explainability facilitates these forms of engagement.

\section{Explainability for Art Practice vs. Art Practice for Explainability}
We draw a distinction between two complementary roles of explainability in the context of AI and art. First, explainability can support artists working with AI by enhancing control, agency, and trust~\cite{tecks_2024_ExplainabilityPathsSustained}. Second, artistic practice can, in turn, contribute to broader societal and cultural understanding of AI—e.g., by raising questions about its production and use~\cite{broad_2024_using, hemment2024experiential}. Explainability can support a behavioural understanding of AI models, i.e. a mental model of how inputs map to outputs. In this work, we argue that explainability is valuable insofar as it helps artists achieve their desired outcomes~\cite{bhattacharya2024towards}, rather than providing a comprehensive account of the model’s inner workings. However, sustained and effective artistic engagement with generative AI often requires some degree of theoretical understanding, such as recognizing the roles of different parameters and components within a generative system, and how these can be manipulated to produce novel outputs that extend beyond the model’s original training or design~\cite{broad_2021_Activedivergencegenerative}. As we will show next, art creation tools that expose and allow the manipulation of different components within a generative AI system can foster both forms of understanding. 

\section{Case Study: Diffusion Model Bending and Inspection in ComfyUI}
ComfyUI is an open-source, node-based interface for designing and executing advanced diffusion pipelines. Unlike other interfaces such as AUTOMATIC1111~\cite{2024_Automatic1111WebUI}, which prioritizes simplicity and abstracts away the model’s internal workings, ComfyUI adopts a low-code, modular approach that decomposes the diffusion process into discrete nodes. These nodes represent stages such as loading base models and applying LoRA adaptations, encoding images into latent representations, generating textual embeddings via the CLIP model, and configuring the denoising process in diffusion. ComfyUI’s backend also manages model loading in a way that helps circumvent GPU memory constraints.

\begin{figure*}[!htb]
    \centering
    \begin{minipage}{0.65\linewidth}
        \centering
        \includegraphics[width=1\linewidth]{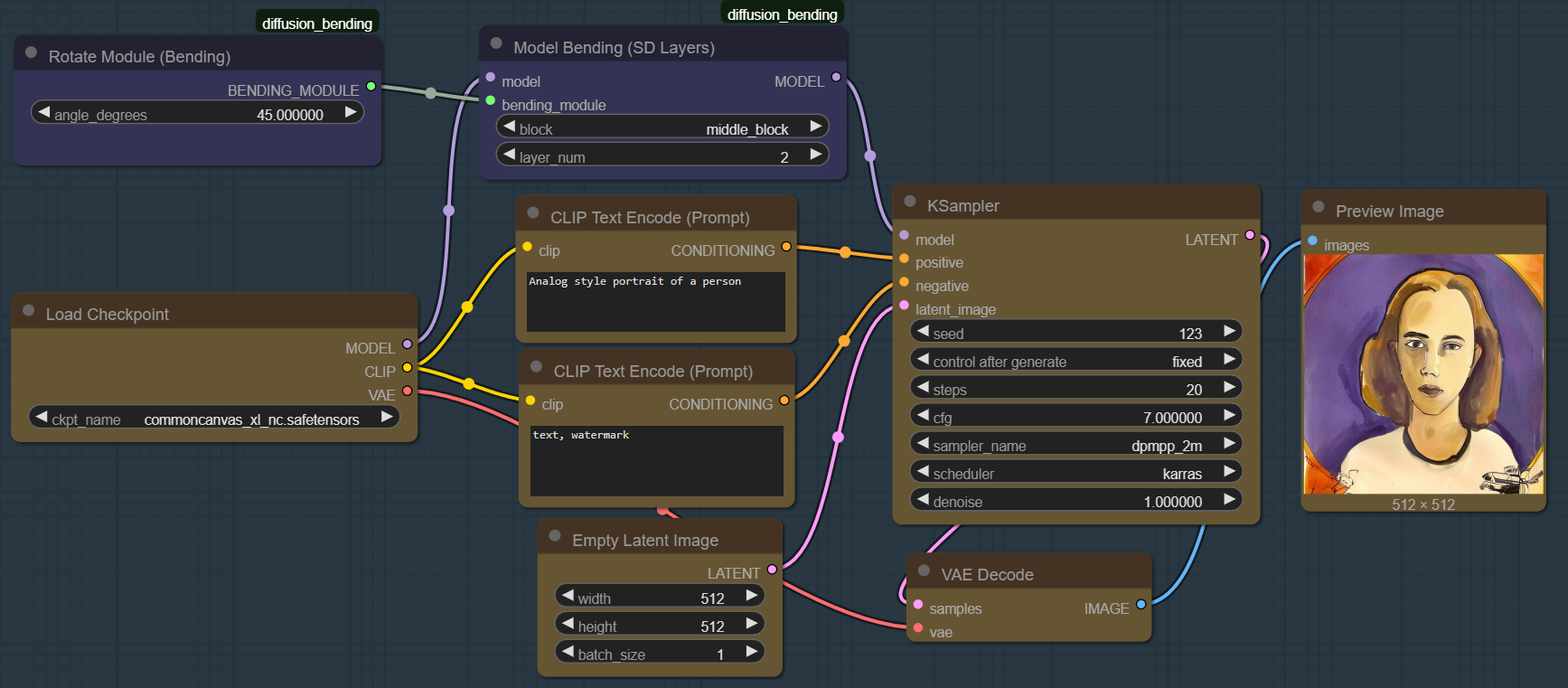}
        \caption{Basic workflow in ComfyUI (gold) running on the open Common Canvas model~\cite{gokaslan2024commoncanvas}. The UNet bending nodes shown in purple.}
        \label{fig:basic_workflow}
    \end{minipage}\hfill
    \begin{minipage}{0.33\linewidth}
        \centering
        \includegraphics[width=\linewidth]{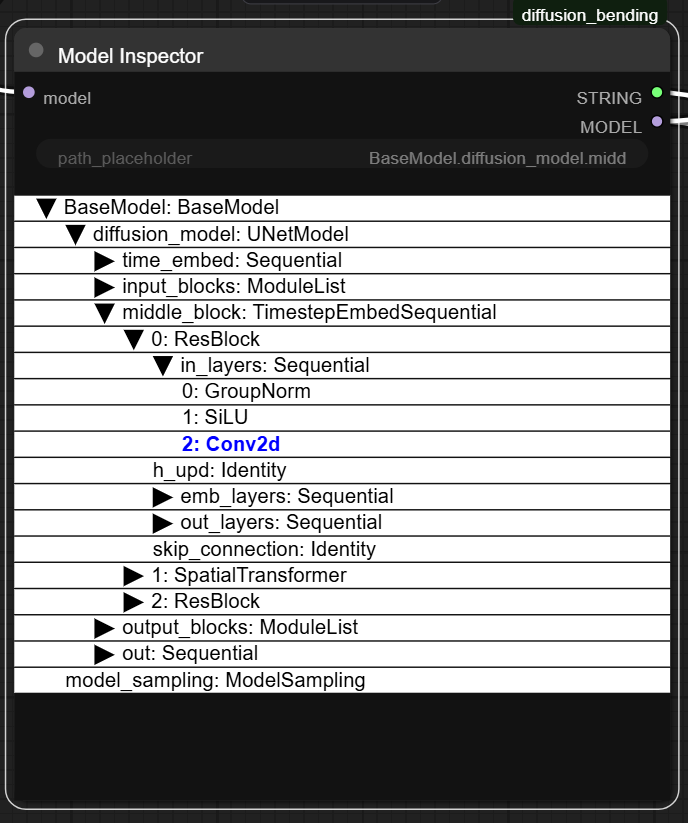}
        \caption{With the model inspector node, artists can interactively pick layers in a model to manipulate. Picked layers are highlighted.}
        \label{fig:model_inspector}
    \end{minipage}
\end{figure*}

We chose ComfyUI because it exposes the inner components of the diffusion pipeline, encouraging users to explore, customize, and develop an understanding of each component. Additionally, ComfyUI frequently integrates state-of-the-art image generation models, making it a timely and flexible testbed for experimentation. A basic workflow in ComfyUI is illustrated in~\autoref{fig:basic_workflow}, and includes steps such as model loading, prompt encoding with CLIP, sampling with user-defined parameters (e.g., number of steps, CFG scale), and decoding the resulting latent image.

\subsection{Related Work}
Network bending is inspired by the tradition of circuit bending—a practice originating in experimental electronic music, in which artists intentionally short-circuit or rewire electronic devices (often children's toys or synthesizers) to produce unpredictable or expressive behaviours. In the context of machine learning, network bending refers to the deliberate manipulation of a model’s internal activations or computational pathways to intervene in its generative process, typically for expressive or exploratory purposes~\cite{broad_2021_NetworkbendingExpressive}. This is achieved by injecting a \textit{bending operator} into a model to transform its intermediate outputs during inference, without the need for additional training or data. Bending operators can perform operations such as adding noise, multiplying/adding scalar values, rotation, and scaling; morphological operations like erosion and dilation, or other custom operations. Previous work on network bending has applied interventions at various layers of the synthesizer network in StyleGAN models~\cite{broad_2021_NetworkbendingExpressive, kraasch_2022_AutolumeLiveTurningGANs, 2024_AutolumeNeuralnetworkbased}. In the context of diffusion models, bending operations have been implemented at user-specified timesteps within the denoising process~\cite{dzwonczyk_2024_NetworkBendingDiffusion} (as inferred from source code), and at select layers of the noise prediction model (UNet)~\cite{curtis_PromptsDevelopingExpressive}.

\subsection{System Design}
In this work, we implement a model bending system as a suite of custom nodes into the ComfyUI interface. The overarching goals of this work are: (1) to provide tools for introducing variations and diversity into the text-to-image diffusion generative process, and (2) these tools should also facilitate a better understanding of the generative process and its parts. The bending operations we implement interject the flow in~\autoref{fig:basic_workflow} by processing the UNet/VAE/LoRA/text-embeddings before passing them downstream. 

In particular, we introduce a set of tools within ComfyUI that enable artists to bend various components of the latent diffusion pipeline, including but not limited to the components covered by previous work~\cite{broad_2021_NetworkbendingExpressive, dzwonczyk_2024_NetworkBendingDiffusion, curtis_PromptsDevelopingExpressive}. These components include the Variational Autoencoder (VAE) used to encode and decode images, the text embeddings produced by the Contrastive Language-Image Pre-training (CLIP) model, the noise prediction process handled by the UNet, as well as at select timesteps during the sampling process. Bending can be used to modify outputs at specific points in the latent diffusion pipeline and to introduce controlled variations, particularly when other generation parameters remain fixed.  Our use of the term bending aligns with the ethos of Broad et al.\cite{broad_2021_NetworkbendingExpressive}, i.e., applying transformations to specific parts of a generative process or model for creative and exploratory purposes. However, it also differs from their original definition in key ways. In our implementation, bending does not exclusively imply injecting custom operators into the model, nor does it require the explicit extraction and manipulation of semantic features. This is in part because large-scale text-to-image diffusion models appear to exhibit semantically rich latent spaces~\footnote{Though certain parts 
(\textit{h-space}) seem to be more potent for semantic manipulation\cite{kwon_2023_DiffusionModelsalready}}. 

The tool can be downloaded manually through a git repository~\footnote{\url{https://github.com/abuzreq/ComfyUI-Model-Bending}} or by installing it through the Comfy Registry~\footnote{\url{https://registry.comfy.org/publishers/abuzreq/nodes/ComfyUI-Model-Bending}}. For brevity, we describe essential parts of the system next and refer those interested to the project's website for details and sample outputs.

\subsection{UNet Model Bending}
Denoising diffusion models recreate images by first adding noise to them in steps, and then learning how to undo that noise. During training, a UNet model learns to clean the noise, so it can later produce new images during inference starting from random inputs~\cite{ho2020denoising}. Given their central role in image generation, UNets are a prime target for bending. UNet models form a U shape with gradually shrinking input blocks that feed to middle blocks, which then expand again to recreate the input images through output blocks. To bend a UNet model, the system requires two components: a bending operator (predefined or custom operators) and a path to the layer to be bent (e.g., \textit{diffusion\_model.middle\_block.0.in\_layers}). For specifying the path to the layer to be bending, users can use the \textit{Model Bending (SD Layers)} node, pick the blocks within the UNet to bend, and a convolution layer within that block (\autoref{fig:basic_workflow}). Once inserted, the bending module becomes part of the model’s inference pipeline, and it is invoked during each denoising step that follows.

\subsection{Model Inspection}
\begin{figure}[ht]
    \centering
    \includegraphics[width=0.8\linewidth]{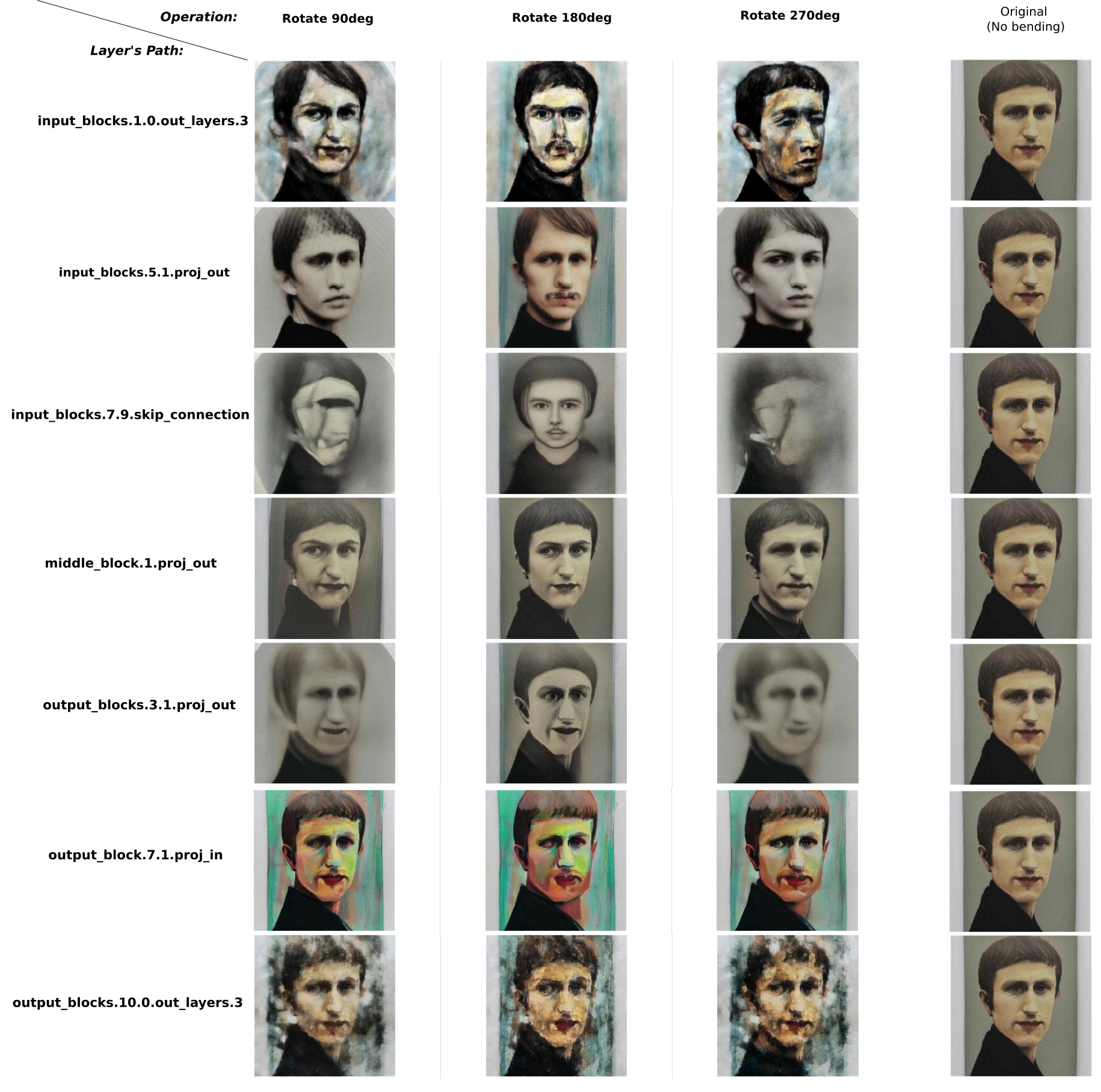}
    \caption{A rotation bending applied to select layers of the SDv1.4 model, prompt "analog style portrait of a person", seed 42, 20 steps, CFG 7, dpmpp\_2m sampler, and Karras scheduler. The result with no bending is shown on the right.}
    \label{fig:rotation_study}
\end{figure}

For more granular control—including the ability to select specific blocks or layers—we provide the Model Inspector node (\autoref{fig:model_inspector}). This tool displays the model’s architecture as a nested, expandable tree, allowing users to visually navigate and select any layer to pass to the model bending node. It currently supports both UNet and VAE models. By interactively bending different parts of the model, artists can develop intuition about how each component influences the output. \autoref{fig:rotation_study} illustrates bending with a rotation operation for select layers in the UNet of the Stable Diffusion v1.4 model. 

\subsection{Feature Maps Visualization}
Feature map visualizations are often used to understand the model's behaviour and the kind of features it is learning at different layers. We provide a Visualize Feature Map node, which allows users to visualize intermediate feature maps at any layer, and the Model Inspector can be used to pick the layers to be visualized interactively. By selecting a layer of interest, artists can inspect what the model is attending to at different stages of the denoising process. These visualizations not only help reveal the evolution of the output but can also be paired with the Model Bending node to observe how specific modifications affect the model’s internal representations.

\subsection{CLIP Text Encoding (Conditionings)}
In a text-to-image system, prompts are encoded into textual embeddings using the CLIP model (referred to as \textit{conditioning}). Our plugin enables users to make fine-grained adjustments within the embedding space, specifically in the local region where the prompt is mapped. Such small movements within the embedding space can help artists understand its structure and offer fine controls that complement prompt-based interaction.

\section{Conclusion and Future Work}
By reframing explainability as an artistic practice grounded in making and manipulation, this paper offers a different lens for thinking about AI-human collaboration. The model bending plugin for ComfyUI invites artists to engage with the inner workings of the large-scale models themselves, treating them as expressive, deformable materials. In the future, we plan to extend this tool with additional forms of model manipulation and custom semantic editing. We also aim to build a community of practice around this tool. Artists might share recipes or suggestions for which parts of a large model yield the most expressive outcomes when bent or their critical reflections while using the tool to inspect models. We hope this approach fosters a deeper, more personal relationship between artists and AI models—one where explainability emerges not just from observation, but from doing. Ultimately, we believe that model crafting tools like the one presented here, which encourage artists to treat large-scale generative models as creative materials, may help address ongoing challenges around authorship and agency in GenAI-assisted art.

\bibliographystyle{ACM-Reference-Format}
\bibliography{main}

\end{document}